\documentclass{aa}
\usepackage{graphicx}
\usepackage{natbib}
\usepackage{txfonts}
\bibpunct{(}{)}{;}{a}{}{,} % to follow the A&A style

\def\ls{LS~5039}

\def\j0632{HESS~J0632$+$057}

\begin{document}

\title{One-dimensional pair cascade emission in gamma-ray binaries}
\subtitle{An upper-limit to cascade emission at superior conjunction in LS 5039}
\titlerunning{One-dimensional pair cascade emission in gamma-ray binaries}

\author{%author1\inst{1,2}
B. Cerutti %\inst{1}
\and G. Dubus %\inst{1}
\and G. Henri %\inst{1}
}
\authorrunning{B. Cerutti et al.}
\institute{
Laboratoire d'Astrophysique de Grenoble, UMR 5571 CNRS, Universit\'e
Joseph Fourier, BP 53, 38041 Grenoble, France 
}

\date{Draft \today}
\abstract
%Context
{In gamma-ray binaries such as \ls\ a large number of electron-positron pairs are created by the annihilation of primary very high energy (VHE) gamma-rays with photons from the massive star. The radiation from these particles contributes to the total high energy gamma-ray flux and can initiate a cascade, decreasing the effective gamma-ray opacity in the system.}
%Aims
{The aim of this paper is to model the  cascade emission and  investigate if it can account for the VHE gamma-ray flux detected by HESS from \ls\ at superior conjunction, where the primary gamma-rays are expected to be fully absorbed.}
%Methods
{A one-dimensional cascade develops along the line-of-sight if the deflections of pairs induced by the surrounding magnetic field can be neglected. A semi-analytical approach can then be adopted, including the effects of the anisotropic seed radiation field from the companion star.}
%Results
{Cascade equations are numerically solved, yielding the density of pairs and photons. In \ls, the cascade contribution to the total flux is large and anti-correlated with the orbital modulation of the primary VHE gamma-rays. The cascade emission dominates close to superior conjunction but  is too strong to be compatible with HESS measurements. Positron annihilation does not produce detectable 511 keV emission.}
%Conclusion
{This study provides an upper limit to cascade emission in gamma-ray binaries at orbital phases where absorption is strong. The pairs are likely to be deflected or isotropized by the ambient magnetic field, which will reduce the resulting emission seen by the observer. Cascade emission remains a viable explanation for the detected gamma-rays at superior conjunction in \ls.
%HESS measurements rule out the presence of an one-dimensional cascade in the system. Yet, this study provides an upper limit from cascades in gamma-ray binaries. Since pair cascading cannot be excluded, a three-dimensional cascade can develop if the magnetic field is strong enough to isotropize pairs.
}

\maketitle

\section{Introduction}

The massive star in gamma-ray binaries plays a key role in the formation of the very high energy (VHE, $>$100~GeV) radiation. The large seed photon density provided by the O or Be companion star, contributes to the production of gamma-rays {\em via} inverse Compton scattering on ultra-relativistic electrons accelerated in the system ({\em e.g.} in a pulsar wind or a jet). The same photons annihilate with gamma-rays, leading to electron-positron pairs production $\gamma+\gamma\rightarrow e^{+}+e^{-}$. In some tight binaries such as \ls, this gamma-ray absorption mechanism is strong if the VHE emission occurs close to the compact object. Gamma-ray absorption can account for the presence of an orbital modulation in the VHE gamma-ray flux from LS 5039, as observed by HESS \citep{2005ApJ...634L..81B,2006MNRAS.368..579B,2006A&A...451....9D}.

%$\gamma\gamma$-absorption is undoubtedly at work in such systems and 
A copious number of pairs may be produced in the surrounding medium as a  by-product of the VHE gamma-ray absorption. If the number of pairs created is large enough and if they have enough time to radiate VHE photons before escaping, a sizeable electromagnetic cascade can be initiated. New generations of pairs and gamma-rays are produced as long 
%as the total energy involved in the pair production interaction is beyond the threshold energy of the reaction. 
as the secondary particles have enough energy to boost stellar photons beyond the  pair production threshold energy. Because of the anisotropic stellar photon field in the system, the inverse Compton radiation produced in the cascade has a strong angular dependence. The cascade contribution depends on the position of the primary gamma-ray source with respect to the massive star and a distant observer.

The VHE modulation in \ls\ was explained in \citet{2008A&A...477..691D} using phase-dependent absorption and inverse Compton emission, ignoring the effect of pair cascading. This model did not predict any flux close to superior conjunction, {\em i.e.} where the massive star lies between the compact object and the observer. This is contradicted by HESS observations \citep{2006A&A...460..743A}. Interestingly, this mismatch intervenes at phases where $\gamma\gamma$-opacity is known to be large $\tau_{\gamma\gamma}\gg 1$. The development of a cascade could contribute to the residual flux observed in the system, with secondary gamma-ray emission filling in for the highly absorbed primary gamma-rays.
%increasing the transparency of the source. 
This possibility was proposed to explain this discrepancy \citep{2006A&A...460..743A} and is quantitatively investigated here in this article.

The ambient magnetic field strength has a critical impact on the development of pair cascading. If the magnetic field strength is small enough to neglect the induced deflections on pair trajectories then the cascade develops along the line of sight joining the primary source of gamma-rays and a distant observer. The particles do not radiate synchrotron radiation.  %In these cases, 
Cascade calculations are then reduced to a one-dimension problem. Such a situation would apply in an unshocked pulsar wind where the pairs are cold relative to the magnetic field carried in the wind. This paper explores the development of an {one-dimensional} pair cascade in a binary and its implications.
%n an unshocked pulsar wind, the magnetic field is frozen into a wind of electron-positron pairs and hence plays no role on the cascade growth.

Previous computations of cascade emission in binary environment were carried out by \citet{1997A&A...322..523B,2005MNRAS.356..711S,2006JPhCS..39..408A,2006MNRAS.368..579B,2007A&A...464..259B,2007A&A...476....9O,2008MNRAS.383..467K,2008APh....30..239S,2009MNRAS.tmpL.175Z}. Except for \citet{2006JPhCS..39..408A}, all these works are based on Monte Carlo methods. One peculiarity of the gamma-ray binary environment is that the source of seed photons for pair production and inverse Compton emission is the high luminosity companion star. This study proposes a semi-analytical model for one-dimensional cascades calculations, taking into account the anisotropy in the seed photon field. The aim of the paper is to investigate and compute the total contribution from pair cascading in the system \ls, and see if it can account for the measured flux close to superior conjunction. The next section presents the main assumptions and equations for cascade computations. The development and the anisotropic effects of pair cascading in compact binaries are investigated. The density of escaping pairs and their rate of annihilation are also calculated in this part. The cascade contribution along the orbit in \ls\ is computed and compared with the available observations in Section~3. The last section concludes on the implications of one-dimensional cascades in gamma-ray binaries. More details about pair production are available in the appendices.%, particularly for \ls. %Comparisons with existing calculations of an unshocked pulsar wind \citep{2008APh....30..239S} are also presented and discussed.

\begin{figure}
\centering
\resizebox{\hsize}{!}{\includegraphics{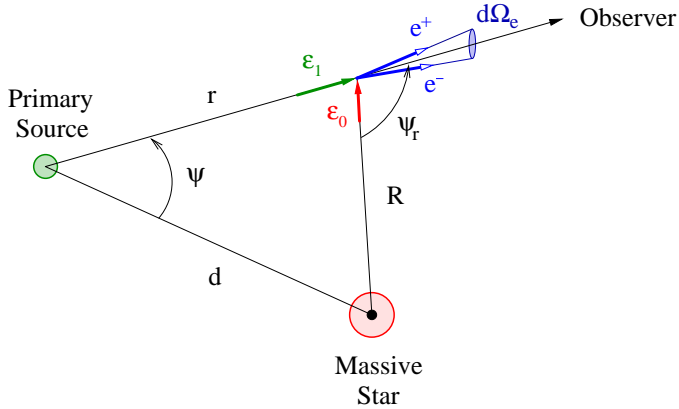}} 
  \caption{This diagram describes the system geometry. A gamma-ray photon of energy $\epsilon_1$ from the primary source (compact object) interacts with a soft photon of energy $\epsilon_0$ at a distance $r$ from the source and $R$ from the massive star (assumed point-like and mono-energetic), producing a pair $e^+/e^-$ boosted toward a distant observer. The system is seen at an angle $\psi$.}
\label{fig_bin}
\end{figure}

\begin{figure}
\centering
\resizebox{\hsize}{!}{\includegraphics{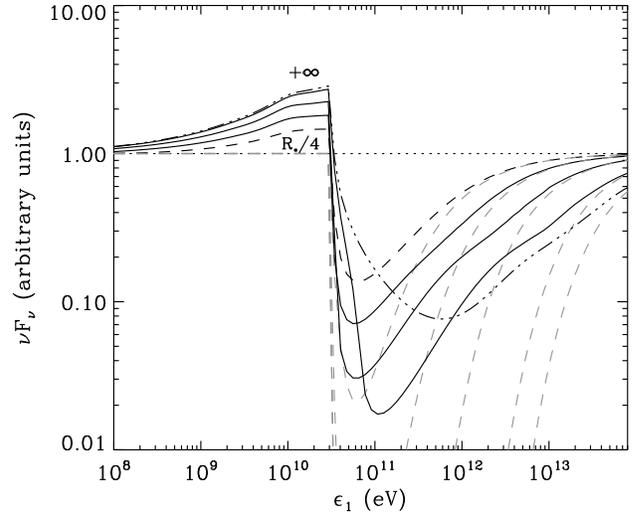}}
  \caption{Cascade development along the path to the observer. The primary source of photons, situated at the location of the compact object, has a power law spectral distribution with photon index -2
%$r\equiv 0$,  injects a $-2$ power distribution 
(dotted line). Spectra are computed using the parameters appropriate for \ls\ at superior conjunction ($d\approx 2 R_{\star}$, $R_{\star}=9.3~R_{\odot}$, $T_{\star}=39~000$ K) for $\psi=30\degr$. The transmitted spectrum, including cascade emission, is shown at various distances from the primary source: $r=R_{\star}/4$ (black dashed line), $R_{\star}/2$, $R_{\star}$, $2 R_{\star}$ (solid lines) and $r=+\infty$ (dotted-dashed line). Pure absorbed spectra are shown for comparison (light dashed line).}
\label{fig_dvp}
\end{figure}

\section{Anisotropic pair cascading in compact binaries}

\begin{figure*}
\resizebox{17cm}{!}
{\includegraphics{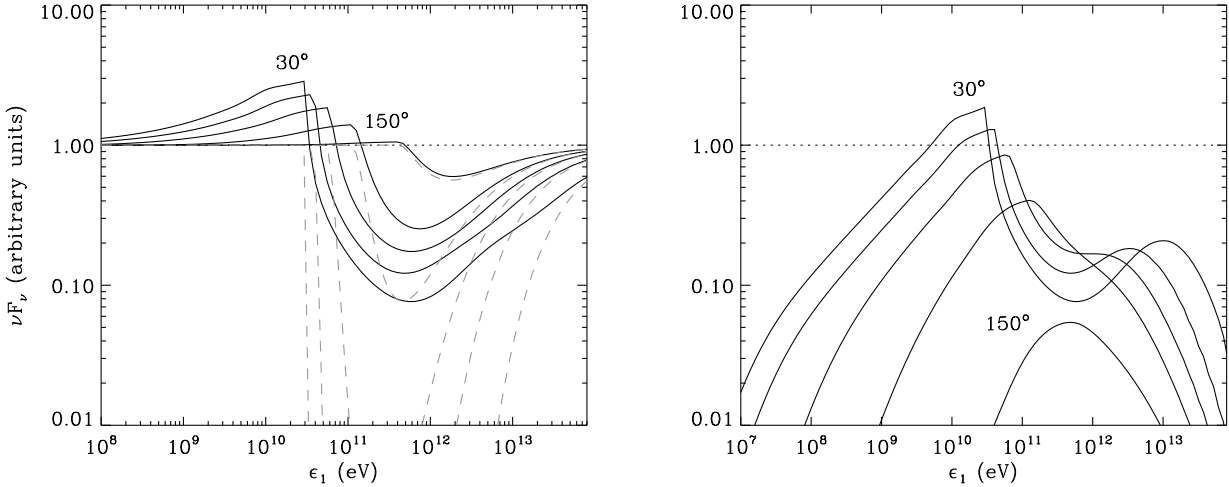}} 
  \caption{Spectra as seen by an observer at infinity, taking into account the effect of cascading. Calculations are applied to \ls\ at periastron for different viewing angle $\psi=30\degr$, $60\degr$, $90\degr$, $120\degr$ and $150\degr$. {\em Left panel}: Complete spectra (solid line) are compared to the pure absorbed (light dashed line) and injected (dotted line) spectra. The contribution from the cascade is presented in the {\em right panel}.}
\label{fig_anis}
\end{figure*}

\subsection{Assumptions}

This part examines one-dimensional cascading in the context of binary systems. The massive star sets the seed photon radiation field for the cascade. For simplicity, the massive star is assumed point-like and mono-energetic. This is a reasonable approximation as previous studies on absorption \citep{2006A&A...451....9D} and emission \citep{2008A&A...477..691D} have shown. The effects of the magnetic field and pair annihilation are neglected (see \S2.5). Triplet pair production (TPP) due to the high energy electrons or positrons propagating in a soft photon field ($\gamma+e^{+-} \rightarrow e^{+-} + e^+ +e^-$, \citealt{1991MNRAS.253..235M}) is not taken into account here. The cross section for this process becomes comparable to inverse Compton scattering when $E_e \epsilon_0\ga 250 (m_e c^2)^2$ that is for electron energies $E_e\ga 6$~TeV interacting with $\epsilon_0\approx 10$ eV stellar photons. With a scattering rate of about $\sim 10^{-2}$ $\rm{s^{-1}}$, only a few pairs can be created {\em via} TPP by each VHE electron, before it escapes or loses its energy in a Compton scattering. The created pairs have much lower energy than the primary electrons. TPP cooling remains inefficient compared to inverse Compton for VHE electrons with energy $\la$~PeV. HESS observations of \ls\ show a break in the spectrum at a few TeV so few electrons are expected to interact by TPP in the cascade. Observations of other gamma-ray binaries also show steep spectra but this assumption will have to be revised if there is significant primary emission beyond $\approx$ 10 TeV. Pair production due to high energy gamma-rays interacting with the surrounding material is also neglected. This occurs for $\gamma$-rays $>$ 1 MeV and the cross-section is of order $0.04 \sigma_T Z^2 ~\rm{cm}^{2}$ (see {\em e.g.} \citealt{1992hea..book.....L}), with $\sigma_T$ the Thomson cross-section. Since the measured $N_H$ is at most $10^{22}$ cm$^{-2}$ in gamma-ray binaries, pair production on matter will not affect the propagation of gamma-rays towards the observer.

Due to the large velocity of the center-of-mass (CM) frame in the observer frame, the direction of propagation of pairs created by $\gamma\gamma$-absorption is boosted in the direction of the initial gamma-ray. For a gamma-ray of energy $\epsilon_1=1~$TeV, the Lorentz factor of the CM to the observer frame transform is $\gamma'\sim \epsilon_1/2 m_e c^2=10^6\gg 1$ (see the appendix, Eq.~\ref{cm_param}). Pairs produced in the cascade are ultra-relativistic with typical Lorentz factor $\gamma_e\sim 10^6\gg 1$. Their emission is forward boosted within a cone of semi-aperture angle $\alpha\sim 1/\gamma_e\ll 1$ in the direction of electrons. The deviations on the electron trajectory due to scattering in the Thomson regime are $\sim \epsilon_0/m_e c^2 \ll 1/\gamma_e$. In the Klein-Nishina regime most of the electron energy is given to the photon. It is assumed here that electrons and photons produced in the cascade remain on the same line, a good approximation since $\gamma'$ and $\gamma_e\gg 1$. This line joins the primary gamma-ray source to a distant observer (Fig.~\ref{fig_bin}).

Pair cascading is one-dimensional as long as magnetic deviations of pairs trajectories along the Compton interaction length $\lambda_{ic}$ remain within the cone of emission of the electrons. This condition holds if $\lambda_{ic}/(2R_L)<1/\gamma_e$, with $R_L$ the Larmor radius. For a typical interaction length $\lambda_{ic}\sim 1/(n_{\star}\sigma_{ic})\sim 10^{11}~$cm for TeV pairs in \ls, the ambient magnetic field must be lower than $B\la 10^{-8}~$G. If the magnetic field strength is much greater, pairs locally isotropize and radiate in all directions. In between, pairs follow the magnetic field lines and the dynamics of each pairs must be followed as treated in \citet{2005MNRAS.356..711S}. The above limit may appear unrealistically stringent. However, since deviations and isotropization will dilute the cascade flux, the one-dimensional approach can be seen as maximising the cascade emission. More exactly, this redistribution induced by magnetic deflections would decrease the cascade flux at orbital phases where many pairs are produced to the benefit of phases where only a few are created. Hence, the one-dimensional approach gives an upper limit to the cascade contribution at phases where absorption is strong. If the flux calculated here using this assumption is lower than required by observations then cascading will be unlikely to play a role. Finally, one-dimensional cascading should hold in the free pulsar wind as long as the pairs move strictly along the magnetic field. In \citet{2005MNRAS.356..711S} and \citet{2008APh....30..239S}, the cascade radiation is computed up to the termination shock using a Monte Carlo approach. \citet{2005MNRAS.356..711S} also include a contribution from the region beyond the shock. The cascade electrons in this region are assumed to follow the magnetic field lines (in contrast with the pulsar wind zone where the propagation is radial). There is no reacceleration at the shock and synchrotron losses are neglected. In the method expounded here, the cascade radiation is calculated semi-analytically from a point-like gamma-ray source at the compact object location up to infinity, providing the maximum possible contribution of the one-dimensional cascade in gamma-ray binaries.

\subsection{Cascade equations}

In order to compute the contribution from the cascade, the radiative transfer equation and the kinetic equation of the pairs have to be solved simultaneously.

The radiative transfer equation for the gamma-ray density $n_{\gamma}\equiv dN_{\gamma}/dt d\epsilon_1 d\Omega$ at a distance $r$ from the source is
\begin{equation}
\frac{dn_{\gamma}}{dr}=-n_{\gamma}\left(\frac{d\tau_{\gamma\gamma}}{dr}\right)+\int n_{\star}\frac{dN}{dtd\epsilon_1}\ n_e\ dE_e,
\label{trans_eq}
\end{equation}
where $n_e\equiv dN_{e}/dr dE_e d\Omega_e$ is the electrons distribution, $n_{\star}$ the seed photon density from the massive star and $dN/dtd\epsilon_1$ the Compton kernel. The kernel is normalised to the soft photon density and depends on the energy $E_e$ of the electron and the angle between the photon and the direction of motion of the electron \citep{2008A&A...477..691D}. In the mono-energetic and point-like star approximation the stellar photon density can be estimated as $L_{\star}/4\pi c R^2 \bar{\epsilon_0}$, where $L_{\star}$ is the stellar luminosity, $\bar{\epsilon_0}\approx 2.7 k T_{\star}$ the mean thermal photon energy and $R$ the distance to the massive star (see Fig.~\ref{fig_bin}). The absorption rate $d\tau_{\gamma\gamma}/dr$ is given by Eq.~(\ref{taukern}), convoluted to the soft photon density.

The kinetic equation for the pairs is given by the following integro-differential equation for $\gamma_e\gg 1$ \citep{1970RvMP...42..237B,1988ApJ...335..786Z,2007A&A...469..857D}
\begin{eqnarray}
\frac{dn_e}{dt} &=&
-n_e(E_e)\int_{m_e c^2}^{E_e}\mathcal{P}\left(E_e,E'_e\right)dE'_e\nonumber\\
&+& \int_{E_e}^{+\infty}n_e(E'_e)\ \mathcal{P}\left(E'_e,E_e\right)dE'_e + 2\int n_{\star}\ g_{\gamma\gamma}\ n_{\gamma}\ d\epsilon_1,
\label{kin_eq}
\end{eqnarray}
where $\mathcal{P}(E_e,E'_e)$ is the transition rate for an electron of energy $E_e$ down-scattered at an energy $E'_e\le E_e$ at $r$. The first two terms on the right side of the equation describe the inverse Compton cooling of pairs, taking into account catastrophic losses in the deep Klein-Nishina regime. In this case, most of the electron energy is lost in the interaction and the scattered photon carries away most of its energy since $\epsilon_1=E_e-E'_e\approx E_e$. A continuous losses equation inadequately describes sizeable stochastic losses in a single interaction \citep{1970RvMP...42..237B,1989ApJ...342.1108Z}.

Since the inverse Compton kernel gives the probability per electron of energy $E_e$ to produce a gamma-ray of energy $\epsilon_1$, the scattering rate can be rewritten as
\begin{equation}
\mathcal{P}(E_e,E_e')=n_{\star}(r)\frac{dN}{dtdE'_e}.
\label{proba-kern}
\end{equation}
The expression of $dN/dtdE'_e$ is the same as the Compton kernel as described before but gives the spectrum of scattered electrons instead of the outcoming photon. The first integral in Eq.~(\ref{kin_eq}) is the inverse Compton scattering rate and can be analytically expressed as
\begin{equation}
\int_{m_e c^2}^{E_e}\mathcal{P}\left(E_e,E'_e\right)dE'_e=\sigma_{ic}\ c\ n_{\star}(r)\left(1-\beta_e\cos\theta_0\right),
\label{rate}
\end{equation}
where $\beta_e$ is the electron velocity in the observer frame and $\sigma_{ic}$ is the total inverse Compton cross-section (for the full expression see {\em e.g.} \citealt{1979rpa..book.....R}, Eq.~7.5). The last term in the kinetic equation is a source of pairs from $\gamma\gamma$-absorption coupled with the photon density (see the appendices). The pair production kernel $g_{\gamma\gamma}$ is normalised to the soft photon density.

The anisotropic cascade can be computed by inserting the anisotropic kernels for inverse Compton scattering (see Eq.~A.6 in \citealt{2008A&A...477..691D}) and for pair production obtained in Eq.~(\ref{kernel}) in Eqs.~(\ref{trans_eq}-\ref{kin_eq}). The following sections present cascade calculations applied to compact binaries, using a simple Runge-Kutta 4 integration method. It is more convenient to perform integrations over an angular variable rather than $r$. Here, calculations are carried out using $\psi_r$, the angle between the line joining the massive star to the observation point and the line of sight (see Fig.~\ref{fig_bin}).

\subsection{Cascade growth along the line of sight}

Figure \ref{fig_dvp} presents cascade calculations for different distances $r$ from the primary gamma-ray source. For illustrative purpose, the source is assumed isotropic and point-like, injecting a power-law distribution of photons with an index $-2$ at $r=0$ but no electrons. The calculations were carried out for a system like \ls\ and for a viewing angle $\psi=30\degr$. In this geometric configuration, absorption is known to be strong ($\tau_{\gamma\gamma}\approx 40$ for 200 GeV photons) and a significant fraction of the total absorbed energy is expected to be reprocessed in the cascade, inverse Compton scattering being also very efficient in this configuration.

Close to the source ($r\la d$ with $d$ the orbital separation), absorption produces a sharp and deep dip in the spectrum (light dashed line) but the cascade starts %contributing 
to fill the gap (black solid line). The angle $\psi_r$ increases with the distance $r$ to the primary source. Hence, the threshold energy for pair production increases as well. Cascading adds more flux to higher energy gamma-rays where absorption is maximum. The cascade produces an excess of low energy gamma-rays below the minimum threshold energy $\epsilon_1\approx 30$ GeV. Because these new photons do not suffer from absorption, they accumulate at lower energies. This is a well-known feature of cascading.

\subsection{Anisotropic effects}

This section investigates anisotropic effects in the development of the cascade as seen by a distant observer. Cascades are computed for different viewing angle $\psi$ at infinity, assuming an isotropic power-law spectrum for the primary gamma-rays.

{\em The left panel} in figure~\ref{fig_anis} shows the complete spectrum taking into account cascading (solid line) compared to the pure absorbed power-law (dashed line). Due to the angular dependence in the pair production process, larger viewing angles shift the cascade contribution to higher energies and decrease its amplitude (Fig.~\ref{fig_anis}, {\em right panel}). The cascade flux is small enough to be ignored for $\psi\ga 150\degr$.

Three different zones can be distinguished in the cascade spectra. First, below the pair production threshold energy, photons accumulate in a low energy tail (photon index $\approx -1.5$) produced by inverse Compton cooling of pairs. For $\psi\la 90\degr$, a low energy cut-off is observed due to the pairs escaping the system \citep{2000APh....12..335B,2008A&A...488...37C}. This low energy cut-off is at about $0.1$~GeV for $\psi=30\degr$. The cutoff occurs when the cascade reaches a distance from the primary source corresponding to $\psi_r\approx 90\degr$.
%because pairs do not have enough time to cool down sufficiently before escaping the system %(the same effect has been noted by \citealt{2000APh....12..335B,2008A&A...488...37C} for an unshocked pulsar wind).
Then, the electrons cannot cool effectively because the inverse Compton interaction angle diminishes and the stellar photon density decreases as they propagate. For $\psi\ga 90\degr$, particles escape right away from the vicinity of the companion star and no tail is produced. Second, above the threshold energy, there is a competition between absorption and gamma-ray production by reprocessed pairs, particularly for small angles where both effects are large. Even if cascading increases the transparency for gamma-rays, absorption still creates a dip in the spectrum. Third, well beyond the threshold energy, absorption becomes inefficient. Fewer pairs are created, producing a high energy cut-off ($\approx 10$ TeV, for $\psi=30\degr$). Klein-Nishina effects also contribute to the decrease of the high energy gamma-rays production.

\subsection{Escaping pairs}

\begin{figure}
\centering
\resizebox{\hsize}{!}{\includegraphics{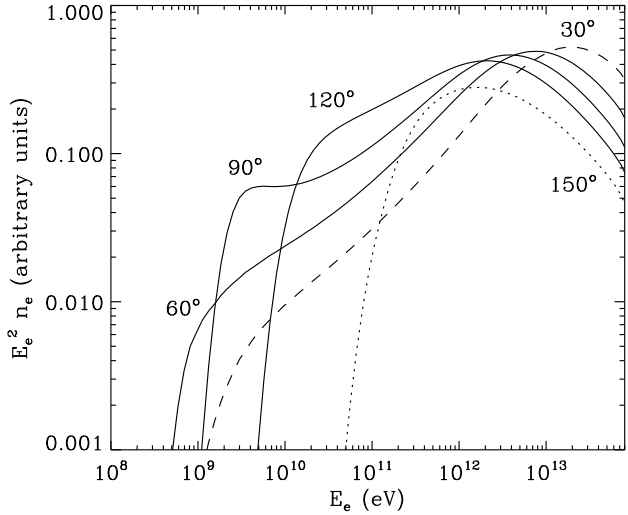}} 
  \caption{Distribution of escaping pairs seen by a distant observer, depending on the viewing angle $\psi=30\degr$ (dashed line), $60\degr$, $90\degr$, $120\degr$ and $150\degr$ (dotted line). The binary parameters are the same than in Fig.~\ref{fig_anis}.}
\label{fig_el}
\end{figure}

The spectrum of pairs produced in the cascade as seen at infinity is shown in figure~\ref{fig_el}. The density depends strongly on the viewing angle as expected, but the mean energy of pairs lies at very high energies ($\left\langle E_e\right\rangle\ga$ 100~GeV, see Table~\ref{tab_comp}). The accumulation of very high energy particles can be explained by two concurrent effects. Far from the massive star ($r\gg d$), most of the pairs are created at very high energy due to the large threshold energy (almost rear-end collision). The second effect is that inverse Compton losses are in deep Klein-Nishina regime for high energy electrons. The cooling timescale increases and becomes larger than the propagation timescale of electrons close to the companion star, producing an accumulation of pairs at very high energies.

The distribution of pairs allows to assess the fraction of the total absorbed energy escaping the system in the form of kinetic energy in the pairs. This non-radiated power $P_{e}$ can be compared to the radiated power released in the cascade $P_{r}$. Energy conservation yields the total absorbed power $P_{a}=P_{e}+P_{r}$.

\begin{table}
\caption{Comparison between the radiated power in the cascade $P_r$ and the absorbed power $P_a$ for different viewing angle $\psi$. This table provides also the mean energy of pairs $\left\langle E_e\right\rangle$ in the cascade at infinity.}
\label{tab_comp}
\centering
\begin{tabular}{c c c c c c}
\hline\hline
\noalign{\smallskip}
$\psi$ & $30\degr$ & $60\degr$ & $90\degr$ & $120\degr$ & $150\degr$ \\
\noalign{\smallskip}
\hline
\noalign{\smallskip}
$P_{r}/P_{a}$ & 80\% & 70\% & 60\% & 40\% & 15\% \\
$\left\langle E_e\right\rangle$ (GeV) & 400 & 100 & 70 & 200 & 1000 \\
%$P_a\times10^{-35}$ (erg/s) & 7.4 & 6.4 & 5.3 & 3.7 & 1.3 \\
%$P_{e}/P_{r}$ & 0.3 & 0.5 & 0.7 & 1.5 & 6 \\
\noalign{\smallskip}
\hline
\end{tabular}
\end{table}

The asymptotic radiated power reached by the cascade is compared to the total absorbed power integrated over energy in Table~\ref{tab_comp}. The fraction of lost energy increases with the viewing angle. In fact, for $\psi>90\degr$ most of the power remains in kinetic energy. Once the electrons are created, only a few have time to radiate through inverse Compton interaction. Below ($\psi<90\degr$), the radiative power dominates and the cascade is very efficient (recycling efficiency up to 80\% for $\psi=30\degr$). The cascade is fully linear, since the power re-radiated remains much smaller than the star luminosity $P_r\ll L_{\star}$ \citep{1987MNRAS.227..403S}. Self-interactions in the cascade are then negligible. This is also a consequence of Klein-Nishina cascading \citep{1988ApJ...335..786Z}. In addition, interactions between particles in the cascade would be forcedly rear-end, hence highly inefficient.

The created positrons will annihilate and form a 511~keV line. However, the expected signal is very weak. The annihilation cross-section is $\sigma\sim \sigma_T \log \gamma / \gamma$ (see {\em e.g.} \citealt{1992hea..book.....L}). The escaping positrons have a very high average Lorentz factor $\gamma \ga 10^5$ (Tab. 1) so they are unlikely to annihilate within the system. They will thermalize and annihilate in the interstellar medium. Escaping positrons from gamma-ray binaries are unlikely to contribute much to the diffuse 511 keV emission. The average number of pairs created along the orbit in \ls\ (based on the results to be discussed in the following section) is $\mathcal{N}_e\sim 5\times 10^{35}~\rm{s}^{-1}$. This estimate does not take into account contributions from triplet pair production or from the pulsar wind (for a pulsar injecting pairs with $\left\langle\gamma_e\right\rangle\sim 10^5$ and a luminosity of $10^{36}~\rm{erg/s}$, about $10^{36}~\rm{s^{-1}}$ pairs are produced). Gamma-ray binaries have short lifetimes and it is unlikely there is more than a few hundred currently active in the Galaxy. Hence, the expected contribution is  orders-of-magnitude below the positron flux required to explain the diffuse 511 keV emission ($\sim 10^{43} ~\rm{s}^{-1}$, \citealt{2005A&A...441..513K}). Even if the positrons thermalize close to or within the system (because magnetic fields contain them, see \S5) then, following \citet{2006A&A...457..753G}, the expected contribution from a single source at 2 kpc would be at most $\sim 10^{-9}$ ph cm$^{-2}$ s$^{-1}$, which is currently well below detectability.

%\begin{figure}
%\centering
%\resizebox{\hsize}{!}{\includegraphics{fig_pow.eps}} 
%   \caption{Reprocessed power by the cascade as a function of $\psi_r$ for $\psi=30\degr$ ({\em top}), $60\degr$, $90\degr$, $120\degr$ and $150\degr$ ({\em bottom}). The same binary parameters were adopted than in Fig.~\ref{fig_anis}.}
%\label{fig_pow}
%\end{figure}

\section{Cascading in \ls}

\begin{figure}
\centering
\resizebox{\hsize}{!}{\includegraphics{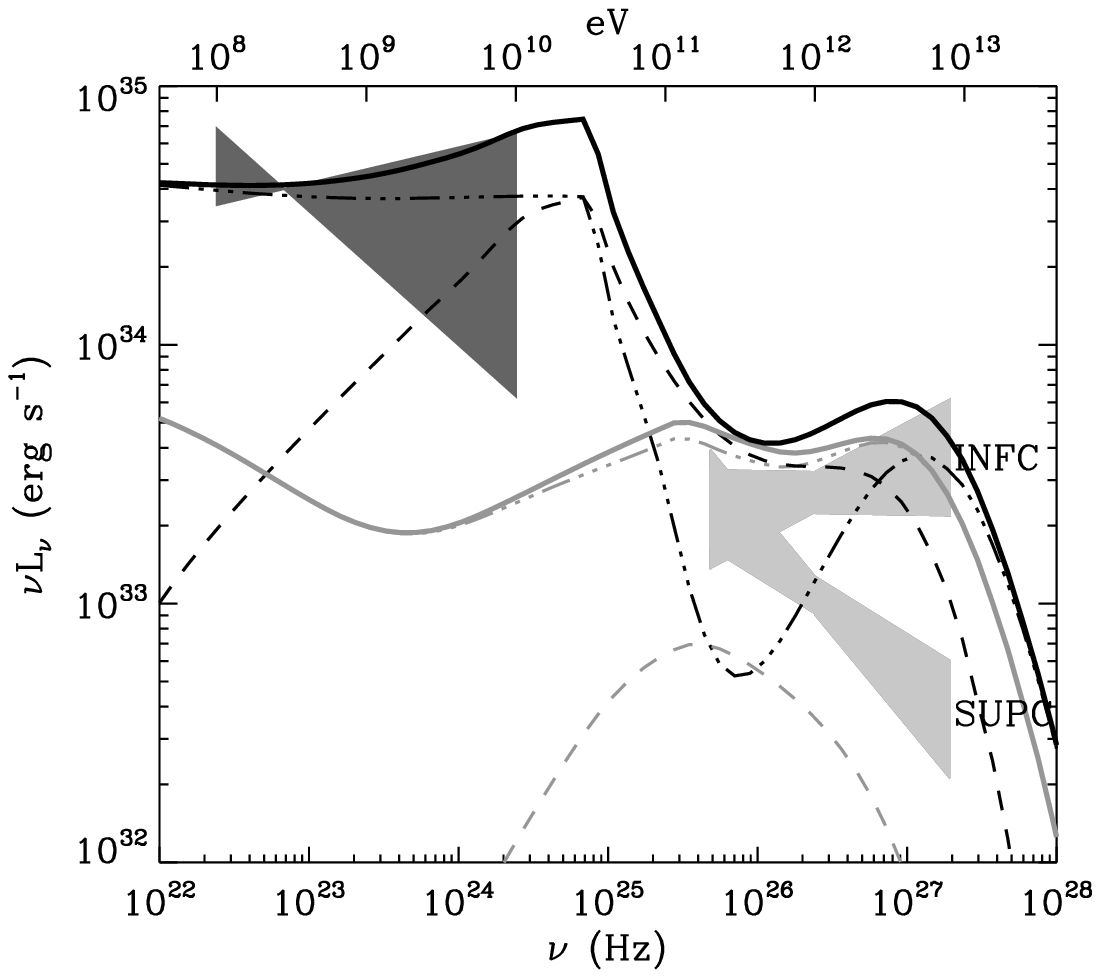}} 
  \caption{Orbit-averaged spectra in \ls\ at INFC ($0.45<\phi<0.9$, grey lines) and SUPC ($\phi<0.45$ or $\phi>0.9$, black lines) and comparisons with EGRET (dark) and HESS (light) bowties \citep{1999ApJS..123...79H,2006A&A...460..743A}. Dotted-dashed lines represent the primary source of gamma-rays with pure absorption, injected at $r\equiv 0$, computed with the model described in \citet{2008A&A...477..691D} for a mono-energetic and point-like star. Dashed lines show the contribution from the cascade and thick solid lines the sum of the primary absorbed source and the cascade contributions.}
\label{fig_cas}
\end{figure}

\begin{figure}
\centering
\resizebox{\hsize}{!}{\includegraphics{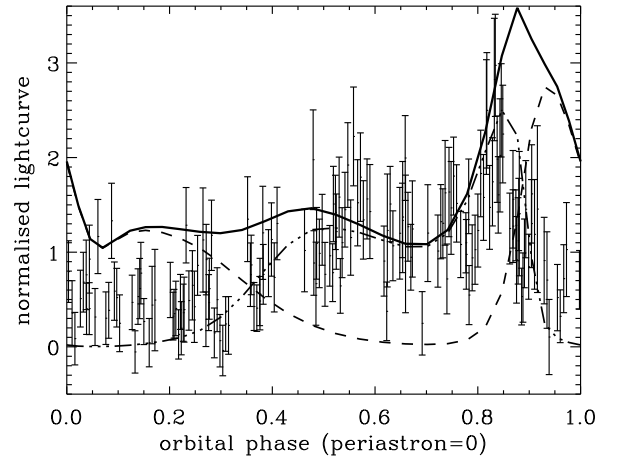}} 
  \caption{Computed light-curves along the orbit in \ls, in the HESS energy band (flux $\ge 100$~GeV). The cascade contribution (dashed line) is compared to the primary pure absorbed source (dotted-dashed line) and HESS observations. The thick solid line shows the sum of both components.}
\label{fig_light}
\end{figure}

%\subsection{Cascade emission along the orbit}

\ls\ was detected by HESS \citep{2005Sci...309..746A} and the orbital modulation of the TeV gamma-ray flux was later on reported in \citet{2006A&A...460..743A}. Most of the temporal and spectral features can be understood as a result of anisotropic gamma-ray absorption and emission from relativistic electrons accelerated in the immediate vicinity of the compact object, {\em e.g.} in the pulsar wind termination shock %a generic one-zone leptonic model 
\citep{2008A&A...477..691D}. 
However, this description fails to explain the residual flux observed close to superior conjunction where a significant excess has been detected (6.1$\sigma$ at phase 0.0$\pm$0.05). The primary gamma-rays should be completely attenuated. The aim of this part is to find if cascading can account for this observed flux. % at superior conjunction.
The cascade is assumed to develop freely from the primary gamma-ray source up to the observer. The contribution of the cascade as a function of the orbital phase is also investigated.

The primary source of gamma-rays now considered is the spectrum calculated in \citet{2008A&A...477..691D}. 
%The primary source is anisotropic and assumed to be produced by a population of relativistic electrons accelerated in the immediate vicinity of the compact object, {\em e.g.} in the pulsar wind termination shock. 
Figure~\ref{fig_cas} shows phase-averaged spectra along the orbit at INFC (orbital phase $0.45<\phi<0.9$) and SUPC ($\phi<0.45$ or $\phi>0.9$) for the primary source, the cascade and the sum of both components. The orbital parameters  and the distance (2.5~kpc) are taken from \citet{2005MNRAS.364..899C} for an inclination $i=60\degr$ so $\psi$ varies between $30\degr-150\degr$. The cascade contribution is highly variable along the orbit and dominates %the complete spectral energy distribution
at SUPC for $\epsilon_1\ga 30$~GeV, where a large pair-production rate is expected. At INFC, cascading is small and negligible compared with the primary flux. With pair cascading the spectral differences between INFC and SUPC are very small at VHE, contrary to what is observed by HESS. In the GeV band, cascades contribute to a spectral hardening at SUPC close to 10-30~GeV.

Orbital light-curves in the HESS energy band give a better appreciation of the contribution from both components (Fig.~\ref{fig_light}). The contribution from cascading is anti-correlated with the primary absorbed flux. The cascade light-curve %peaks a bit further {\bf from} the superior conjunction ($\phi\approx 0.9$) and 
is minimum at inferior conjunction ($\phi\approx 0.72$). The non trivial double peaked structure of the lightcurve at phases 0.85-0.35 is due to competition in the cascade between absorption and inverse Compton emission.
% both maximum at this phase, producing this non trivial double peaked structure.
Absorption has a slight edge at superior conjunction  ($\phi\approx 0.06$), producing a dip at this phase.
%A secondary dip {\bf occurs} %is predicted 
%at superior conjunction ($\phi\approx 0.06$) owing to the competition between absorption and inverse Compton emission both maximum at this phase, producing this non trivial double peaked structure. 
Elsewhere, the primary contribution dominates over the cascade emission. %cascade can be neglected and %the modulation can then be mainly explained by anisotropic inverse Compton production light-curve (see Fig.~3, {\em top panel} in 
%\citealt{2008A&A...477..691D}). 
At lower energies ($\epsilon_1<10~$GeV), the cascade contribution is undistinguishable from the primary source.

In this configuration, the cascade does add VHE gamma-ray emission close to superior conjunction but the expected contribution overestimates HESS observations. Decreasing the inclination of the system does not help: %makes things even worse. Indeed, 
the cascade flux in the TeV energy band increases, since the primary source is on average more absorbed along the orbit (see \S3 in \citealt{2006A&A...451....9D}). For $i\la 30\degr$, the cascade contribution dominates the primary flux at every orbital phases in the VHE band. % Another possibility would be to change the primary particle distribution injected in the pulsar wind along the orbit, but then a new model for the primary source would have to be formulated.
One-dimension cascades can be ruled out by the current HESS observations of \ls.

%\subsection{Cascading in an unshocked pulsar wind}
%\subsection{Comparison with a Monte-Carlo approach}

%If most of the very high energy radiation originates from an unshocked pulsar wind, as computed in \citet{2008A&A...488...37C,2008arXiv0811.2466S,2008APh....30..239S} for \ls\ and \lsi, one-dimensional cascading calculations are appropriate. In this case, the magnetic field plays no role since it is frozen into a radially expanding wind of pairs. Pair cascading develops along the line of sight pulsar-observer. Most of the observed high energy radiation is assumed to originate from  the pulsar wind zone (PWZ) where it is produced by the primary pairs and by the cascade \citep{2008APh....30..239S}. The pair cascade is followed only up to the termination shock, which reduces the cascade contribution compared to the case described above where the cascade continues to infinity.
%Radiation from the cascade is neglected 
%Any other contribution to the emission beyond the termination shock is also ignored. The semi-analytical method expounded here was applied to the PWZ model and compared to the Monte Carlo computations. Similar results were obtained in the same configurations for a mono-energetic and power-law injection of pairs at the location of the pulsar in \ls.

\section{Conclusion}

This paper explored the impact of one-dimensional pair cascading on the formation of the very high energy radiation from gamma-ray binaries in general, \ls\ specifically. A significant fraction of the total absorbed energy can be reprocessed at lower energy by the cascade, decreasing the global opacity of the primary source. Anisotropic effects also play a major role on the cascade radiation spectrum seen by a distant observer.

A large contribution from cascading is expected in \ls, large enough that it significantly overestimates the flux observed by HESS. %Assuming a lower inclination for the system makes things even worse, since more pairs are created on average. 
%Available observations from HESS rule out one-dimension cascading in \ls. 
One-dimensional cascading is too efficient in redistributing the absorbed primary flux and can be ruled out. However, the fact that it overestimates the observed flux means a more general cascade cannot be ruled out (it would have been if the HESS flux had been underestimated).
%In particular, such cascade cannot explain the residual flux detected at superior conjunction as it was initially intended for. 
%However, this study may provide an upper limit on pair cascading in gamma-ray binaires. 
%This would be particularly true close to superior conjunction where most of the cascade takes place, since all the radiation is boosted along the line of sight. As a result, pair cascading cannot be excluded in \ls. 
%However, this does not rule out a more general pair cascade. 
If the ambient magnetic field is large enough ($B\gg 10^{-8}~$G) the pairs will be deflected from the line-of-sight. For $B\ga 10^{-3}$ G the Larmor radius of a TeV electron becomes smaller than the \ls\ orbital separation and the pairs will be more and more isotropised locally. All of this will tend to dilute cascade emission compared to the one-dimensional case, which should therefore be seen as an upper limit to the cascade contribution at orbital phases where absorption is strong, particularly at superior conjunction. The initiated cascade will be three-dimensional as pointed out by \citet{1997A&A...322..523B}. Each point in the binary system becomes a potential secondary source able to contribute to the total gamma-ray flux at every orbital phases. Cascade emission can still be sizeable all along the orbit in \ls, yet form a more weakly modulated background in the light-curve on account of the cascade radiation redistribution at other phases. %If pairs cools preferentially via synchrotron radiation rather than by inverse Compton scattering, the cascade will be quenched as noticed in \citet{2008MNRAS.383..467K}.
The strength and structure of the surrounding magnetic field (from both stars) has a strong influence on the cascade \citep{2005MNRAS.356..711S,2008A&A...482..397B,2008A&A...489L..21B}. More realistic pair cascading calculations cannot be treated with the semi-analytical approach exposed here. Complementary investigations using a Monte Carlo approach are needed to better appreciate the cascade contribution in gamma-ray binaries.

Finally, the cascade will be quenched if the created pairs lose energy to synchrotron rather than inverse Compton scattering. This requires ambient magnetic fields $B \ga 5$ G, as found by equating the radiative timescales for a 1 TeV electron at periastron in 
\ls. Such ambient magnetic field strengths could be reached close to the companion star. In this case an alternative explanation is needed to account for the flux at superior conjunction. A natural one to consider is that the primary gamma-ray source is more distant to the massive star. The VHE source would not be coincident with the compact object location anymore and would suffer less from absorption. In the microquasar scenario, \citet{2007A&A...464..259B} can account for consistent flux with HESS observations at superior conjunction if some electrons are injected well above the orbital plane (jet altitude $z>10~R_{\star}$). In addition to \ls, this possibility was also considered for the system Cyg~X$-$1 by \citet{2008A&A...489L..21B} and \citet{2009MNRAS.tmpL.175Z}.

In practice, reality may consist of a complex three-dimensional cascade partly diluted and partly quenched depending upon position, angle and magnetic field configuration.

\begin{acknowledgements}
GD thanks A. Mastichiadis for discussions on triplet pair production. This work was supported by the {\em European Community} via contract ERC-StG-200911.
\end{acknowledgements}

%\onecolumn
\appendix

\section{Pair production}

The main equations for the pair production process are briefly presented here. Detailed calculations can be found in \citet{1967PhRv..155.1408G}, \citet{1971MNRAS.152...21B} and \citet{1997A&A...325..866B}.

\subsection{Kinematics and cross-sections}

The interaction of a gamma-ray photon of energy $\epsilon_1$ and a soft photon of energy $\epsilon_0$ in the observer frame leads to the production of an electron-positron pair if the total available energy in the center-of-mass (CM) frame is greater than the rest mass energy of the pair
\begin{equation}
2\epsilon_1\epsilon_0\left(1-\cos\theta_0\right)\ge 4m_e^2 c^4,
\label{kin}
\end{equation}
where $m_e$ is the electron mass and $\theta_0$ the angle between the two incoming photons in the observer frame. It is useful to define the Lorentz invariant $s=\epsilon_1\epsilon_0\left(1-\cos\theta_0\right)/2$. Pairs are produced if $s\ge m_e^2 c^4$ and the velocity $\beta$ of the electron-positron pair in the CM frame is $\beta=(1-m_e^2 c^4/s)^{1/2}$.

The differential cross-section $d\sigma_{\gamma\gamma}/d(\beta\cos\theta'_1)$ in the CM frame depends on $\beta$ and the angle $\theta'_1$ between the outcoming electron-positron pair and the incoming photons. The full expression can be found in {\em e.g.} \citet{1971MNRAS.152...21B}, Eq.~(2.7). The differential cross-section presents a symmetric structure, peaked at $\cos\theta'_1=\pm 1$ and minimum for $\cos\theta'_1=0$. Electrons are mostly created in the same and opposite direction with respect to the incoming hard photon direction in the CM frame. The double peaked structure is enhanced with increasing energy ($s\gg m_e^2 c^4$) and becomes less pronounced close to the threshold ($s\sim m_e^2 c^4$). The integration over the angles gives the total pair production cross-section $\sigma_{\gamma\gamma}$, maximum close to the threshold (see Eq.~1 in \citealt{1967PhRv..155.1408G}).

The construction of the CM frame with respect to the observer frame can be simplified if one of the incoming photons carries most of the energy. This case is appropriate in the present context. For $\epsilon_1\gg\epsilon_0$, the CM frame can be considered as propagating along the same direction as the high energy photon. The velocity of the CM frame in the observer frame can be expressed as
\begin{equation}
\beta'=\left(1-\frac{4s}{\epsilon^2_1}\right)^{1/2}.
\label{cm_param}
\end{equation}
The total energy of say the electron $E_e$ in the observer frame can then be formulated using the Lorentz transform from the CM to the observer frames
\begin{equation}
E_e=\gamma'\left[s^{1/2}+\beta'\left(s-m_e^2 c^4\right)^{1/2}\cos\theta'_1\right],
\label{en_obs}
\end{equation}
providing a relation between $E_e$ and $\cos\theta'_1$.

\subsection{Rate of absorption and pair spectrum kernels}

A gamma-ray photon going through a soft photon gas of density $dn/d\epsilon d\Omega$ is absorbed at a rate per unit of path length $l$
\begin{equation}
\frac{d\tau_{\gamma\gamma}}{dl}=\iint\frac{dn}{d\epsilon d\Omega} \left(1-\cos\theta\right)\sigma_{\gamma\gamma}d\epsilon d\Omega.
\label{dtau}
\end{equation}
The absorption rate gives the probability for a gamma-ray of energy $\epsilon_1$ to be absorbed but does not give the energy of the pair created in the interaction.

Following \citet{1971MNRAS.152...21B}, the probability for a gamma-ray of energy $\epsilon_1$ to be absorbed between $l$ and $l+dl$ yielding an electron of energy between $E_e$ and $E_e+dE_e$ (with a positron of energy $E_{e^+}\approx\epsilon_1-E_e$ for $\epsilon_1\gg\epsilon$) is
\begin{equation}
g_{\gamma\gamma}=\iint\frac{dn}{d\epsilon d\Omega} \left(1-\cos\theta\right)\frac{d\sigma_{\gamma\gamma}}{dE_e}d\epsilon d\Omega.
\label{gkern}
\end{equation}
As with anisotropic inverse Compton scattering \citep{2008A&A...477..691D}, it is useful to consider the case of a monoenergetic beam of soft photons. The normalised soft photon density in the observer frame is
\begin{equation}
\frac{dn}{d\epsilon d\Omega}=\delta\left(\epsilon-\epsilon_0\right)\delta\left(\cos\theta-\cos\theta_0\right)\delta\left(\phi-\phi_0\right),
\label{dnph}
\end{equation}
where $\delta$ is the Dirac distribution. Injecting Eq.~(\ref{dnph}) into Eq.~(\ref{gkern}) gives the anisotropic pair production kernel, a convenient tool for spectral computations. The detailed calculation is presented in Appendix~B and the complete expression given in Eq.~(\ref{kernel}). The pair production kernel has a strong angular dependence and a symmetric structure, centered at $E_e=\epsilon_1/2$ and peaked at $E_e=E_{\pm}$ (see Appendix~B, Fig.~\ref{kern_gg}). The effect of the angle $\theta_0$ is reduced close to the threshold where the particles share equally the energy of the primary gamma-ray photon $E_e\approx E_{e^+}\approx\epsilon_1/2$. Far from the threshold, one particle carries away almost all the available energy $E_e\approx\epsilon_1$.

The anisotropic kernel integrated over all the pitch angles, in the case of an isotropic gas of photons, is consistent with the kernel found by \citet{1983Afz....19..323A}. Note that a general expression for the anisotropic kernel valid beyond the approximation $\epsilon_1\gg\epsilon_0$ is presented in \citet{1997A&A...325..866B}.

\subsection{Pair density}

The number of pair created per unit of length path and electron energy depends on the probability to create a pair and on the probability for the incoming gamma-ray to remain unabsorbed up to the point of observation so that
\begin{equation}
\frac{dN_e}{dl dE_e}=\left\{g_{\gamma\gamma}\left(E_e\right)+g_{\gamma\gamma}\left(\epsilon_1-E_e\right)\right\}e^{-\tau_{\gamma\gamma}(l)}.
\label{edens}
\end{equation}
Because of the symmetry in $g_{\gamma\gamma}$ and since electrons and positrons cannot be distinguished here, $g_{\gamma\gamma}\left(\epsilon_1-E_e\right)=g_{\gamma\gamma}\left(E_e\right)$. The integration over electron energy yields
\begin{equation}
\frac{dN_e}{dl}=2\left(\int g_{\gamma\gamma}\left(E_e\right)dE_e\right)e^{-\tau_{\gamma\gamma}(l)}=2\frac{d\tau_{\gamma\gamma}}{dl}e^{-\tau_{\gamma\gamma}(l)}.
\label{dnedl}
\end{equation}
The total number of pairs produced by a single gamma-ray bathed in a soft radiation along the path $l$ up to the distance $r$ is then
\begin{equation}
N_e(r)=2\left(1-e^{-\tau_{\gamma\gamma}(r)}\right).
\label{ner}
\end{equation}
For low opacity $\tau_{\gamma\gamma}\ll 1$, pair production is inefficient and the number of particles produced tends to $\approx 2\tau_{\gamma\gamma}$. For high opacity $\tau_{\gamma\gamma}\gg 1$, a pair is always created.

\section{Anisotropic pair production kernel}

\begin{figure}
\centering
\resizebox{\hsize}{!}{\includegraphics{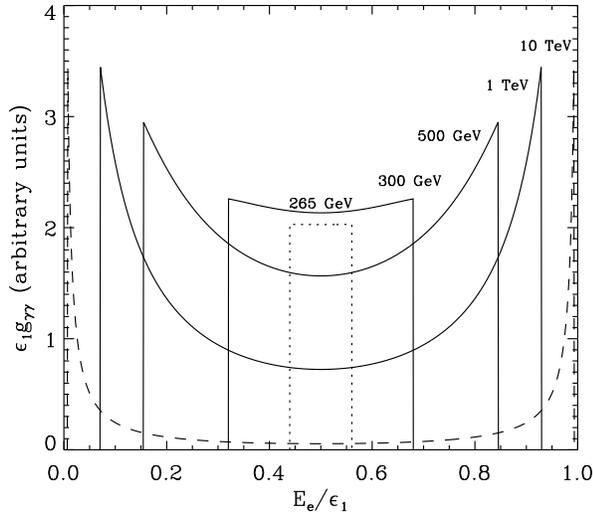}} 
  \caption{Anisotropic pair production kernel $g_{\gamma\gamma}$ with $\epsilon_0$ set at 1~eV for a head-on collision ($\theta_0=\pi$). The kernel is computed for $\epsilon_1=265$~GeV (dotted line), $300$~GeV, $500$~GeV, $1$~TeV and $10$~TeV (dashed line). The yielding of pairs occurs for $\epsilon_1\ge 260$~GeV.}
\label{kern_gg}
\end{figure}

This section is dedicated to the calculation of the pair energy spectrum produced in the interaction between a single gamma-ray photon of energy $\epsilon_1$ and a mono-energetic beam of soft photons. The general expression in Eq.~(\ref{gkern}) can be reformulated using the relativistic invariant $s$
\begin{equation}
g_{\gamma\gamma}=\frac{4}{\epsilon^2_1}\iiint \frac{s}{\epsilon^2_0}\frac{dn}{d\epsilon d\Omega}\frac{d\sigma_{\gamma\gamma}}{dE_e} d\epsilon ds d\phi.
\label{gskern}
\end{equation}
Combining the expression of $\beta$ with the equations Eqs.~(\ref{cm_param}-\ref{en_obs}) and defining $x\equiv\gamma'^2$, the differential cross-section variables can be written as
\begin{equation}
\beta(x)=\left(1-\frac{4 m_e^2 c^4 x}{\epsilon^2_1}\right)^{1/2}, \\
\beta\cos\theta'_1(x)=\frac{2E_e-\epsilon_1}{\epsilon_1\left(1-\frac{1}{x}\right)^{1/2}}.
\label{betcos}
\end{equation}
The differential cross-section can then be expressed as
\begin{eqnarray}
\frac{d\sigma_{\gamma\gamma}}{dE_e} &=& \frac{d\sigma_{\gamma\gamma}}{d\left(\beta\cos\theta'_1\right)}\frac{d\left(\beta\cos\theta'_1\right)}{dE_e} \nonumber\\
&=&\frac{2}{\epsilon_1\left(1-\frac{1}{x}\right)^{1/2}}\frac{d\sigma_{\gamma\gamma}}{d\left(\beta\cos\theta'_1\right)}.
\label{dsig}
\end{eqnarray}
The complete general formula to compute the spectrum of the pair for a non-specified soft radiation field is
\begin{equation}
g_{\gamma\gamma}=\frac{\epsilon_1}{4}\iiint\frac{1}{\epsilon^2 x^3}\frac{2}{\left(1-\frac{1}{x}\right)^{1/2}}\frac{dn}{d\epsilon d\Omega}\frac{d\sigma_{\gamma\gamma}}{d(\beta\cos\theta'_1)}d\epsilon dx d\phi,
\label{bonrees}
\end{equation}
corresponding to Eq.~(2.14) in \citet{1971MNRAS.152...21B}. The injection of a mono-energetic and unidirectional soft photon density (Eq.~\ref{dnph}) in this last equation yields
\begin{equation}
g_{\gamma\gamma}=\frac{2\left(1-\mu_0\right)}{\epsilon_1 \left(1-\frac{1}{x_0}\right)^{1/2}}\frac{d\sigma_{\gamma\gamma}}{d(\beta\cos\theta'_1)}\left\{\beta\left(x_0\right),\beta\cos\theta'_1\left(x_0\right)\right\},
\label{kernel}
\end{equation}
where $\mu_0\equiv\cos\theta_0$ and
\begin{equation}
x_0=\frac{\epsilon_1}{2\epsilon_0\left(1-\mu_0\right)}.
\label{x0}
\end{equation}
This expression is valid for $\epsilon_1\gg\epsilon_0$ and $s\ge m_e^2 c^4$. The minimum $E_-$ and maximum $E_+$ energy reached by the particles is set by the kinematics of the reaction and given by
\begin{equation}
E_{\pm}=\frac{\epsilon_1}{2}\left[1\pm\left(1-\frac{1}{x_0}\right)^{1/2}\left(1-\frac{4 m_e^2 c^4 x_0}{\epsilon^2_1}\right)^{1/2}\right].
\label{eminmax}
\end{equation}
Figure~\ref{kern_gg} presents the pair production kernel for different incoming gamma-ray energy $\epsilon_1$.

Note that a kernel can be calculated as well for the absorption rate. Injecting Eq.~(\ref{dnph}) into Eq.~(\ref{dtau}) is straightforward and gives
\begin{equation}
\frac{d\tau_{\gamma\gamma}}{dl}=\left(1-\cos\theta_0\right)\sigma_{\gamma\gamma}\left(\beta\right).
\label{taukern}
\end{equation}

\bibliographystyle{aa}
\bibliography{caslin}

\begin{thebibliography}{34}
\expandafter\ifx\csname natexlab\endcsname\relax\def\natexlab#1{#1}\fi

\bibitem[{{Aharonian} {et~al.}(2005){Aharonian}, {Akhperjanian}, {Aye},
  {Bazer-Bachi}, {Beilicke}, {Benbow}, {Berge}, {Berghaus}, {Bernl{\"o}hr},
  {Boisson}, {Bolz}, {Borrel}, {Braun}, {Breitling}, {Brown}, {Gordo},
  {Chadwick}, {Chounet}, {Cornils}, {Costamante}, {Degrange}, {Dickinson},
  {Djannati-Ata{\"i}}, {Drury}, {Dubus}, {Emmanoulopoulos}, {Espigat},
  {Feinstein}, {Fleury}, {Fontaine}, {Fuchs}, {Funk}, {Gallant}, {Giebels},
  {Gillessen}, {Glicenstein}, {Goret}, {Hadjichristidis}, {Hauser},
  {Heinzelmann}, {Henri}, {Hermann}, {Hinton}, {Hofmann}, {Holleran}, {Horns},
  {Jacholkowska}, {de Jager}, {Kh{\'e}lifi}, {Komin}, {Konopelko}, {Latham},
  {Le Gallou}, {Lemi{\`e}re}, {Lemoine-Goumard}, {Leroy}, {Lohse}, {Marcowith},
  {Martin}, {Martineau-Huynh}, {Masterson}, {McComb}, {de Naurois}, {Nolan},
  {Noutsos}, {Orford}, {Osborne}, {Ouchrif}, {Panter}, {Pelletier}, {Pita},
  {P{\"u}hlhofer}, {Punch}, {Raubenheimer}, {Raue}, {Raux}, {Rayner}, {Reimer},
  {Reimer}, {Ripken}, {Rob}, {Rolland}, {Rowell}, {Sahakian}, {Saug{\'e}},
  {Schlenker}, {Schlickeiser}, {Schuster}, {Schwanke}, {Siewert}, {Sol},
  {Spangler}, {Steenkamp}, {Stegmann}, {Tavernet}, {Terrier}, {Th{\'e}oret},
  {Tluczykont}, {Vasileiadis}, {Venter}, {Vincent}, {V{\"o}lk}, \&
  {Wagner}}]{2005Sci...309..746A}
{Aharonian}, F., {Akhperjanian}, A.~G., {Aye}, K.-M., {et~al.} 2005, Science,
  309, 746

\bibitem[{{Aharonian} {et~al.}(2006{\natexlab{a}}){Aharonian}, {Akhperjanian},
  {Bazer-Bachi}, {Beilicke}, {Benbow}, {Berge}, {Bernl{\"o}hr}, {Boisson},
  {Bolz}, {Borrel}, {Braun}, {Brown}, {B{\"u}hler}, {B{\"u}sching}, {Carrigan},
  {Chadwick}, {Chounet}, {Cornils}, {Costamante}, {Degrange}, {Dickinson},
  {Djannati-Ata{\"i}}, {O'C.~Drury}, {Dubus}, {Egberts}, {Emmanoulopoulos},
  {Espigat}, {Feinstein}, {Ferrero}, {Fiasson}, {Fontaine}, {Funk}, {Funk},
  {F{\"u}{\ss}ling}, {Gallant}, {Giebels}, {Glicenstein}, {Goret},
  {Hadjichristidis}, {Hauser}, {Hauser}, {Heinzelmann}, {Henri}, {Hermann},
  {Hinton}, {Hoffmann}, {Hofmann}, {Holleran}, {Horns}, {Jacholkowska}, {de
  Jager}, {Kendziorra}, {Kh{\'e}lifi}, {Komin}, {Konopelko}, {Kosack},
  {Latham}, {Le Gallou}, {Lemi{\`e}re}, {Lemoine-Goumard}, {Lohse}, {Martin},
  {Martineau-Huynh}, {Marcowith}, {Masterson}, {Maurin}, {McComb}, {Moulin},
  {de Naurois}, {Nedbal}, {Nolan}, {Noutsos}, {Orford}, {Osborne}, {Ouchrif},
  {Panter}, {Pelletier}, {Pita}, {P{\"u}hlhofer}, {Punch}, {Raubenheimer},
  {Raue}, {Rayner}, {Reimer}, {Reimer}, {Ripken}, {Rob}, {Rolland}, {Rowell},
  {Sahakian}, {Santangelo}, {Saug{\'e}}, {Schlenker}, {Schlickeiser},
  {Schr{\"o}der}, {Schwanke}, {Schwarzburg}, {Shalchi}, {Sol}, {Spangler},
  {Spanier}, {Steenkamp}, {Stegmann}, {Superina}, {Tavernet}, {Terrier},
  {Tluczykont}, {van Eldik}, {Vasileiadis}, {Venter}, {Vincent}, {V{\"o}lk},
  {Wagner}, \& {Ward}}]{2006A&A...460..743A}
{Aharonian}, F., {Akhperjanian}, A.~G., {Bazer-Bachi}, A.~R., {et~al.}
  2006{\natexlab{a}}, \aap, 460, 743

\bibitem[{{Aharonian} {et~al.}(2006{\natexlab{b}}){Aharonian}, {Anchordoqui},
  {Khangulyan}, \& {Montaruli}}]{2006JPhCS..39..408A}
{Aharonian}, F., {Anchordoqui}, L., {Khangulyan}, D., \& {Montaruli}, T.
  2006{\natexlab{b}}, Journal of Physics Conference Series, 39, 408

\bibitem[{{Aharonian} {et~al.}(1983){Aharonian}, {Atoian}, \&
  {Nagapetian}}]{1983Afz....19..323A}
{Aharonian}, F.~A., {Atoian}, A.~M., \& {Nagapetian}, A.~M. 1983, Astrofizika,
  19, 323

\bibitem[{{Ball} \& {Kirk}(2000)}]{2000APh....12..335B}
{Ball}, L. \& {Kirk}, J.~G. 2000, Astroparticle Physics, 12, 335

\bibitem[{{Bednarek}(1997)}]{1997A&A...322..523B}
{Bednarek}, W. 1997, \aap, 322, 523

\bibitem[{{Bednarek}(2006)}]{2006MNRAS.368..579B}
{Bednarek}, W. 2006, \mnras, 368, 579

\bibitem[{{Bednarek}(2007)}]{2007A&A...464..259B}
{Bednarek}, W. 2007, \aap, 464, 259

\bibitem[{{Blumenthal} \& {Gould}(1970)}]{1970RvMP...42..237B}
{Blumenthal}, G.~R. \& {Gould}, R.~J. 1970, Reviews of Modern Physics, 42, 237

\bibitem[{{Bonometto} \& {Rees}(1971)}]{1971MNRAS.152...21B}
{Bonometto}, S. \& {Rees}, M.~J. 1971, \mnras, 152, 21

\bibitem[{{Bosch-Ramon} {et~al.}(2008{\natexlab{a}}){Bosch-Ramon},
  {Khangulyan}, \& {Aharonian}}]{2008A&A...482..397B}
{Bosch-Ramon}, V., {Khangulyan}, D., \& {Aharonian}, F.~A. 2008{\natexlab{a}},
  \aap, 482, 397

\bibitem[{{Bosch-Ramon} {et~al.}(2008{\natexlab{b}}){Bosch-Ramon},
  {Khangulyan}, \& {Aharonian}}]{2008A&A...489L..21B}
{Bosch-Ramon}, V., {Khangulyan}, D., \& {Aharonian}, F.~A. 2008{\natexlab{b}},
  \aap, 489, L21

\bibitem[{{B{\"o}ttcher} \& {Dermer}(2005)}]{2005ApJ...634L..81B}
{B{\"o}ttcher}, M. \& {Dermer}, C.~D. 2005, \apjl, 634, L81

\bibitem[{{B\"{o}ttcher} \& {Schlickeiser}(1997)}]{1997A&A...325..866B}
{B\"{o}ttcher}, M. \& {Schlickeiser}, R. 1997, \aap, 325, 866

\bibitem[{{Casares} {et~al.}(2005){Casares}, {Rib{\'o}}, {Ribas}, {Paredes},
  {Mart{\'{\i}}}, \& {Herrero}}]{2005MNRAS.364..899C}
{Casares}, J., {Rib{\'o}}, M., {Ribas}, I., {et~al.} 2005, \mnras, 364, 899

\bibitem[{{Cerutti} {et~al.}(2008){Cerutti}, {Dubus}, \&
  {Henri}}]{2008A&A...488...37C}
{Cerutti}, B., {Dubus}, G., \& {Henri}, G. 2008, \aap, 488, 37

\bibitem[{{D'Avezac} {et~al.}(2007){D'Avezac}, {Dubus}, \&
  {Giebels}}]{2007A&A...469..857D}
{D'Avezac}, P., {Dubus}, G., \& {Giebels}, B. 2007, \aap, 469, 857

\bibitem[{{Dubus}(2006)}]{2006A&A...451....9D}
{Dubus}, G. 2006, \aap, 451, 9

\bibitem[{{Dubus} {et~al.}(2008){Dubus}, {Cerutti}, \&
  {Henri}}]{2008A&A...477..691D}
{Dubus}, G., {Cerutti}, B., \& {Henri}, G. 2008, \aap, 477, 691

\bibitem[{{Gould} \& {Schr{\'e}der}(1967)}]{1967PhRv..155.1408G}
{Gould}, R.~J. \& {Schr{\'e}der}, G.~P. 1967, Physical Review, 155, 1408

\bibitem[{{Guessoum} {et~al.}(2006){Guessoum}, {Jean}, \&
  {Prantzos}}]{2006A&A...457..753G}
{Guessoum}, N., {Jean}, P., \& {Prantzos}, N. 2006, \aap, 457, 753

\bibitem[{{Hartman} {et~al.}(1999){Hartman}, {Bertsch}, {Bloom}, {Chen},
  {Deines-Jones}, {Esposito}, {Fichtel}, {Friedlander}, {Hunter}, {McDonald},
  {Sreekumar}, {Thompson}, {Jones}, {Lin}, {Michelson}, {Nolan}, {Tompkins},
  {Kanbach}, {Mayer-Hasselwander}, {M{\"u}cke}, {Pohl}, {Reimer}, {Kniffen},
  {Schneid}, {von Montigny}, {Mukherjee}, \& {Dingus}}]{1999ApJS..123...79H}
{Hartman}, R.~C., {Bertsch}, D.~L., {Bloom}, S.~D., {et~al.} 1999, \apjs, 123,
  79

\bibitem[{{Khangulyan} {et~al.}(2008){Khangulyan}, {Aharonian}, \&
  {Bosch-Ramon}}]{2008MNRAS.383..467K}
{Khangulyan}, D., {Aharonian}, F., \& {Bosch-Ramon}, V. 2008, \mnras, 383, 467

\bibitem[{{Kn{\"o}dlseder} {et~al.}(2005){Kn{\"o}dlseder}, {Jean}, {Lonjou},
  {Weidenspointner}, {Guessoum}, {Gillard}, {Skinner}, {von Ballmoos},
  {Vedrenne}, {Roques}, {Schanne}, {Teegarden}, {Sch{\"o}nfelder}, \&
  {Winkler}}]{2005A&A...441..513K}
{Kn{\"o}dlseder}, J., {Jean}, P., {Lonjou}, V., {et~al.} 2005, \aap, 441, 513

\bibitem[{{Longair}(1992)}]{1992hea..book.....L}
{Longair}, M.~S. 1992, {High energy astrophysics. Vol.1: Particles, photons and
  their detection}, ed. M.~S. {Longair}

\bibitem[{{Mastichiadis}(1991)}]{1991MNRAS.253..235M}
{Mastichiadis}, A. 1991, \mnras, 253, 235

\bibitem[{{Orellana} {et~al.}(2007){Orellana}, {Bordas}, {Bosch-Ramon},
  {Romero}, \& {Paredes}}]{2007A&A...476....9O}
{Orellana}, M., {Bordas}, P., {Bosch-Ramon}, V., {Romero}, G.~E., \& {Paredes},
  J.~M. 2007, \aap, 476, 9

\bibitem[{{Rybicki} \& {Lightman}(1979)}]{1979rpa..book.....R}
{Rybicki}, G.~B. \& {Lightman}, A.~P. 1979, {Radiative processes in
  astrophysics} (New York, Wiley-Interscience, 1979.~393 p.)

\bibitem[{{Sierpowska} \& {Bednarek}(2005)}]{2005MNRAS.356..711S}
{Sierpowska}, A. \& {Bednarek}, W. 2005, \mnras, 356, 711

\bibitem[{{Sierpowska-Bartosik} \& {Torres}(2008)}]{2008APh....30..239S}
{Sierpowska-Bartosik}, A. \& {Torres}, D.~F. 2008, Astroparticle Physics, 30,
  239

\bibitem[{{Svensson}(1987)}]{1987MNRAS.227..403S}
{Svensson}, R. 1987, \mnras, 227, 403

\bibitem[{{Zdziarski}(1988)}]{1988ApJ...335..786Z}
{Zdziarski}, A.~A. 1988, \apj, 335, 786

\bibitem[{{Zdziarski}(1989)}]{1989ApJ...342.1108Z}
{Zdziarski}, A.~A. 1989, \apj, 342, 1108

\bibitem[{{Zdziarski} {et~al.}(2009){Zdziarski}, {Malzac}, \&
  {Bednarek}}]{2009MNRAS.tmpL.175Z}
{Zdziarski}, A.~A., {Malzac}, J., \& {Bednarek}, W. 2009, \mnras, L175+

\end{thebibliography}

\end{document}